\documentclass[final,authoryear,12pt]{elsarticle}

\usepackage{amssymb}

\usepackage{color}

\usepackage{pslatex}
\usepackage{graphicx}
\usepackage{lineno}
\newcommand{\apj}{Astrophys. J.}
\newcommand{\jgr}{J. Geophys. Res.}
\newcommand{\grl}{Geophys. Res. Lett.}
\newcommand{\planss}{Planet. Space Sci.}
\newcommand{\ssr}{Space Sci. Rev.}
\newcommand{\cotwo}{\mbox{CO$_{2}$}}
\newcommand{\htwoo}{\mbox{H$_{2}$O}}
\newcommand{\lya}{{Lyman-$\alpha$}}
\newcommand{\lyb}{{Lyman-$\beta$}}
\newcommand{\arcdeg}{\mbox{$^\circ$}}%
\newcommand{\rmars}{{$R_{\rm M}$}}
\newcommand{\rosetta}{{\it Rosetta}}

\def\Hone{H\,{\sc i}}

\def\Cone{C\,{\sc i}}

\def\Oone{O\,{\sc i}}

\def\lam{$\lambda$}

\journal{Icarus}

\begin{document}

\begin{frontmatter}



\title{Rosetta-Alice Observations of Exospheric Hydrogen and Oxygen on Mars}


\author[jhu]{Paul D. Feldman\corref{cor1}}
\cortext[cor1]{Corresponding author}
\ead{pdf@pha.jhu.edu}

\author[swb]{Andrew J. Steffl}

\author[swb]{Joel Wm. Parker}

\author[umd]{Michael F. A'Hearn}

\author[sa]{Jean-Loup Bertaux}

\author[swb]{S. Alan Stern}

\author[apl]{Harold A. Weaver}

\author[swsa]{David C. Slater}

\author[swsa]{Maarten Versteeg}

\author[swb]{Henry B. Throop}

\author[nwu]{Nathaniel J. Cunningham}

\author[umd]{Lori M. Feaga}

\address[jhu]{Johns Hopkins University, Department of Physics and
Astronomy, 3100 N. Charles Street, Baltimore, MD 21218 USA}

\address[swb]{Southwest Research Institute, 
Department of Space Studies, Suite 300, 
1050 Walnut Street, 
Boulder CO 80302-5150 USA}

\address[umd]{Department of Astronomy, 
University of Maryland, 
College Park MD 20742-2421 USA}

\address[sa]{LATMOS, CNRS/UVSQ/IPSL, 11 Boulevard d'Alembert, 78280 
Guyancourt, France}

\address[apl]{Johns Hopkins University Applied Physics Laboratory,
Space Department,
11100 Johns Hopkins Road,
Laurel, MD 20723-6099 USA}

\address[swsa]{Southwest Research Institute, P. O. Drawer 28510, 
San Antonio TX 78228-0510 USA}

\address[nwu]{Physics Department, Nebraska Wesleyan University, 
Lincoln, NE 68504-2794 USA}

\begin{abstract}

The European Space Agency's \rosetta\ spacecraft, en route to a 2014
encounter with comet 67P/Churyumov-Gerasimenko, made a gravity assist
swing-by of Mars on 25 February 2007, closest approach being at 01:54
UT.  The Alice instrument on board \rosetta, a lightweight
far-ultraviolet imaging spectrograph optimized for {\it in situ}
cometary spectroscopy in the 750--2000 \AA\ spectral band, was used to
study the daytime Mars upper atmosphere including emissions from
exospheric hydrogen and oxygen.  Offset pointing, obtained five hours
before closest approach, enabled us to detect and map the HI \lya\ and
\lyb\ emissions from exospheric hydrogen out beyond 30,000 km from the
planet's center.  These data are fit with a Chamberlain exospheric
model from which we derive the hydrogen density at the 200~km exobase
and the H escape flux.  The results are comparable to those found from
the the Ultraviolet Spectrometer experiment on the {\it Mariner 6} and {\it 7}
fly-bys of Mars in 1969.  Atomic oxygen emission at 1304~\AA\ is
detected at altitudes of 400 to 1000~km above the limb during limb
scans shortly after closest approach.  However, the derived oxygen
scale height is not consistent with recent models of oxygen escape
based on the production of suprathermal oxygen atoms by the
dissociative recombination of O$_2^+$.

\end{abstract}


\end{frontmatter}


\section{Introduction}
\label{intro}

The extended atomic hydrogen corona of Mars was first detected by the
ultraviolet spectrometer experiments on the {\it Mariner~6} and {\it 7}
spacecraft that measured resonantly scattered solar \lya\ radiation
\citep{Barth:1971} to a planetocentric distance of 24,000~km.
This was followed by a similar experiment on the orbiting {\it Mariner 9}
mission \citep{Barth:1972}.  \citet{Anderson:1971}, using radiative
transfer theory, analyzed the early data to derive a hydrogen escape
rate, which they found to be compatible with the water
photodissociation rate at Mars.  \citet{Barth:1972} recognized that the
apparently constant H escape rate derived from {\it Mariner 6, 7}, and
{\it 9}, if extended backward over geological time scales, would result
in an oxygen abundance in the lower atmosphere several orders of
magnitude larger than observed.  \citet{McElroy:1972} suggested that
dissociative recombination of O$_2^+$, the dominant ion in the
atmosphere, would produce oxygen atoms with sufficient energy to
escape.  However, since the energetic oxygen atoms are mostly produced
below the exobase, determination of the escaping fraction requires
detailed modeling, represented by the recent work of
\citet{Lammer:2003}, \citet{Fox:2009}, \citet{Shematovich:2007},
\citet{Valeille:2009a,Valeille:2010a}, and others.  
\citet{Barth:1971} also reported observations of \Oone\ \lam 1304 emission
up to 700~km.  These data, together with those from {\it Mariner~9}, were
interpreted by \citet{Strickland:1973} in terms of a cool, optically
thick oxygen exosphere.  Since then, the only measurement of exospheric
oxygen on Mars is from the SPICAM instrument on {\it Mars Express}
\citep{Chaufray:2009}, but those data extended only up to 400 km where
radiative transfer effects in the \Oone~\lam1304 multiplet are
significant and the evidence for a hot component of O atoms was
inconclusive.

Until recently there have also been no additional measurements of the
extended hydrogen atmosphere on Mars.  \citet{Chaufray:2008}, again
with SPICAM, have measured \Hone~\lya\ emission at 1216~\AA, but
only up to 4,000 km.  Again, they find that a two temperature H
exosphere is possible, although the result is inconclusive.
\citet{Clarke:2009} also suggest a two-component H distribution from
 monochromatic \lya\ images of Mars showing the H corona out to 4 Mars
radii (\rmars) taken by the Solar Blind Channel of the Advanced Camera
for Surveys on {\it HST}.  Radiative transfer modeling is necessary to
extract densities from both of these data sets.
\citeauthor{Clarke:2009} also noted a variation in the \lya\ brightness
over a period of two months which they ascribe to a seasonal variation
in the \htwoo\ loss rate.

We report here on observations of both the extended hydrogen and oxygen
coronae of Mars made with the Alice far-ultraviolet imaging
spectrograph \citep{Stern:2007} on \rosetta\ during the spacecraft's
gravity assist swing-by of Mars on 25 February 2007.  Offset exposures
enabled us to detect and map the \Hone~\lya\ and \lyb\ emissions to
beyond 30,000 km from the planet's center.  Moreover, except near the
planet's limb, the \lyb\ emission is optically thin, allowing us to
use a spherical Chamberlain model to determine the temperature and
density of H at the exobase without the need for radiative transfer
modeling.  From limb pointings at spacecraft distances closer to the
planet, oxygen emission above the exobase is detected, allowing us to
constrain the density of hot oxygen without the need for radiative transfer
modeling.  These data can be used to derive atomic escape rates and
address the question of stoichiometric loss of water vapor from Mars.

\section{Observations}
\label{obs}
\rosetta\ approached Mars from the day side making its closest approach (CA)
at 01:54 UT on 25 February 2007 at an altitude of 250~km.  In order to
satisfy the scientific goals of the various remote sensing instruments
\citep[see, e.g.,][]{Coradini:2010}, the common instrument boresight was
programmed for a number of fixed pointings towards both the sunlit and
dark hemispheres of Mars and offset from Mars, as well as raster scans
across the sunlit limb.  Due to operational constraints, observations
were not possible during the immediate CA period.  The pointings of
interest in this paper (denoted by the operational designation ALxx)
were AL03 pre-CA, centered on the sunlit disk, AL10E, offset pointings
centered 2.5\arcdeg\ and 7.5\arcdeg\ from Mars along the equator, and
AL11B, scans across the illuminated crescent post-CA.  Observation
start times and geometry parameters are given in Table~\ref{obstable}.
At the time of closest approach, Mars was 1.445~AU from the Sun and
the areocentric longitude, $L_s$, was 189.9\arcdeg.  Solar activity was
very low for an extended time, including when Earth faced the same solar
longitude 9 days earlier, with $F_{10.7} \approx$72 at 1~AU.

Alice is a lightweight, low-power, imaging spectrograph optimized for
{\it in situ} cometary far-ultraviolet (FUV) spectroscopy. It is
designed to obtain spatially-resolved spectra in the
750-2000~\AA\ spectral band with a spectral resolution between 8 and
and 12~\AA\ for extended sources that fill its field-of-view.  The slit
is in the shape of a dog bone, 5.5\arcdeg\ long, with a width of
0.05\arcdeg\ in the central 2.0\arcdeg\ while the ends are
0.10\arcdeg\ wide.  Each spatial pixel along the slit is 0.30\arcdeg.  Alice
employs an off-axis telescope feeding a 0.15-m normal incidence Rowland
circle spectrograph with a concave holographic reflection grating. The
imaging microchannel plate detector utilizes dual solar-blind opaque
photocathodes (KBr and CsI) and employs a two-dimensional delay-line
readout.  Details of the instrument are given by \citet{Stern:2007}.

The Alice slit geometry, illustrating the shape of the multi-segment
slit, is shown in Fig.~\ref{offset} for the first pre-CA offset
pointing of 2.5\arcdeg.  The second offset moved the slit an additional
5.0\arcdeg\ away from Mars parallel to the Martian equator.  Except
near the limb, the only features seen in the offset spectra are
\Hone~\lya\ and \lyb, and these data are used to extract the spatial
profiles of these emissions.

A similar diagram for the post-CA limb scans, beginning 25 February
2007 at UT 03:33:02, is shown in Fig.~\ref{limb}.  At this time each
spatial pixel projected to 280~km in altitude but as \rosetta\ receded
from Mars the projected size of each pixel increased.  The scan slowly
shifted the boresight $\sim$250~km towards Mars over a 15-minute
period.  This is illustrated in Fig.~\ref{limb_time}.  Because of the
scanning motion, these spectra were acquired in ``pixel-list'' mode,
that is the position of each photon count is recorded together with a
time tag so that spectra could be reconstructed with the motion
accounted for.  Because the \rosetta-Alice instrument has only a single
data buffer, the data gaps seen in Fig.~\ref{limb_time} result from the
time required to read out the buffer to the spacecraft.  The typical
time to fill the buffer was 30~s, so in practice we accumulated
individual spectra corresponding to $\sim$30~s integrations.  Even so,
to detect \Oone\ emission at high altitudes, we need to co-add multiple
spectra, as described below.

\section{Data Analysis}
\label{data}

\subsection{Calibration}
\label{cal}

During the gravity assist swing-bys of both Mars and the Earth, the
\rosetta\ instruments were powered on and operated primarily to provide
flight verification of instrument performance and to acquire calibration
data such as standard ultraviolet star fluxes and detector flat-fields.  
There were also opportunities to exercise the full range of instrument
parameters that could be adjusted by remote command in flight in order
to optimize the signal-to-noise performance of the instrument.  For the
Mars swing-by, the detector high voltage level was set at --3.8~kV.
Subsequent operations and analysis showed the optimum setting to be
--3.9~kV and all observations beginning in the fall of 2007 were made
at that voltage.  Nevertheless, by comparison of stellar standards at
different voltages (and at different times), we are able to transfer
the current absolute flux calibration to the epoch of the Mars swing-by
and this is incorporated into the current version of the data pipeline
(version 3) with which all of the data have since been processed.  The
data used in this study are publicly available, both from NASA's
Planetary Data System ({\tt http://pdssbn.astro.umd.edu/}) and 
ESA's Planetary Science Archive 
({\tt http://www.rssd.esa.int/?project=PSA}).

To derive spatial profiles of the observed emissions requires an
accurate flat-field calibration.  This poses a rather acute problem
for \lya\ at 1216~\AA\ because the instrument was designed for this
wavelength to fall in the gap between the KBr and CsI photocathode
coatings.  The intent was to utilize the low detection efficiency of
the bare microchannel plate to compensate for the very high expected \lya\
photon flux from the comet.  However, because of a slight misalignment
in the coating edges, the \lya\ sensitivity varies by a factor of two
along the portion of the detector that is mapped onto the sky.  In contrast,
at \lyb\ the variation across the slit is $\sim$1.4.  There is also a
$\pm$20\% odd-even detector row effect that is accentuated at the lower
detector high voltage of --3.8~kV.

Fortuitously, following the Mars encounter, \rosetta-Alice was pointed
towards Jupiter and recorded many hours of spectra in support of the
{\it New Horizons} fly-by of Jupiter in February 2007.  These
observations were made with the same instrument parameters used for the
Mars observations.  Since, from the orbit of Mars the Jovian system
only filled a single row of the detector, high S/N measurements of
detector flat-fields at \lya\ and \lyb\ were obtained, together with a
measure of grating scattered \lya\ as a function of detector row.  An
added bonus is that since Jupiter was only 20\arcdeg\ away on the sky
from the coordinates of the offset pointing, these observations
provided a measure of the interplanetary \lya\ and \lyb\ background for
the Mars observations.  We used a co-added accumulation of 630,000~s of
data obtained between 1 March and 10 March 2007 to derive the
flat-fields used in the analysis described below.

\subsection{Mars dayglow spectrum}
\label{dayglow}

We briefly discuss the dayglow spectrum of Mars, obtained under
AL03 pre-CA.  It consisted of four exposures, each of 1028 seconds. A
composite of the five central rows of the sum of four exposures is
shown in Fig.~\ref{day_spec}.  The viewing parameters are given in
Table~\ref{obstable}.  At the start of the sequence, the angular
diameter of Mars was 1.62\arcdeg\ so that the five central rows of the
detector were uniformly filled with the illuminated Martian disk.  For
comparison with previous work, we also show the Mars full disk spectrum
recorded by the Hopkins Ultraviolet Telescope (HUT) on board the
Space Shuttle in March 1995 (a
full solar cycle earlier)  \citep{Feldman:2000}, convolved to the
spectral resolution of Alice.  At the time, Mars was 1.666~AU from the
Sun, $L_s$ was 70.5\arcdeg, and $F_{10.7}$ was $\approx$75.  The HUT
spectrum is multiplied by a factor of 0.80 to match the brightness
observed by Alice, and this is quite good agreement considering the
Alice calibration uncertainty and the fact that the HUT spectrum
measured the integrated disk brightness, not just a central stripe of
the disk.  The comparison also serves to validate the wavelength
calibration and provides a reference spectrum with which to compare the
exospheric spectra obtained in AL10E and AL11B.  As noted by
\citet{Barth:1972} and \citet{Leblanc:2006}, with the exception of the
\Hone\ Lyman series and \Oone\ \lam 1304, Mars' dayglow emissions,
principally CO and \Cone, are confined to altitudes below
$\sim$200~km.

\subsection{Offset exposures}
\label{offset_exp}

The two offset exposures were taken immediately following the full
disk spectra as \rosetta\ approached Mars.  \Hone\ \lyb\ was detected along
the full length of the Alice slit for both offsets.  Because the slit is
segmented, for each row along the slit the observed \lyb\ signal was fit
to a gaussian profile superimposed on a grating-scattered background that
was fit to a second-order polynomial.  The flux was then obtained by
integrating under the gaussian and then correcting for the detector
flat-field that was derived from subsequent Jupiter observations as
described in Section~\ref{cal}.  The result is shown as a histogram in
Fig.~\ref{lb_offset}.  The errors shown are statistical in the count
rate.  The \lya\ profile is similarly derived except that the detector
background, due to dark counts, was negligible, and is shown in
Fig.~\ref{la_offset}.  The shape of this profile is very similar to the
slant intensity profiles derived from the {\it Mariner 6} and {\it 7} fly-bys by
\citet{Barth:1971}, although lower in absolute brightness as the {\it Mariner}
fly-bys occurred at a time of high solar activity.  Neither 
Fig.~\ref{lb_offset} nor Fig.~\ref{la_offset} have had the interplanetary
background subtracted as did the plots of \citeauthor{Barth:1971}  The
model fits to the data are discussed below in Section~\ref{hmodel}.

\subsection{Limb scan spectra}
\label{limbscan}

From the data acquired in pixel list mode during the limb scans
schematically illustrated in Fig.~\ref{limb_time}, we can extract a
spectrum spanning a given time interval.  However, because the line-of-sight
of the spectrogram was changing its position above the limb with time,
it is necessary to balance the motion with the need for a sufficiently
long integration time to obtain an adequate signal-to-noise ratio for
weak emission features.  At the same time, we need to avoid contamination
of the spectrum by thermospheric emissions.  This was done by co-adding
the photon counts from the first four exposures for row 14; the first 12
exposures for row 15; and the last 14 exposures for rows 16 and 17.  This
results in an altitude weighted average with a trapezoidal shape of
$\approx$320~km centered at 420, 665, 910, and 1240~km, respectively
for rows 14 to 17, respectively.  

Examples of extracted spectra for rows 15 and 16 are shown in
Fig.~\ref{limbspec}.  Note that only \Hone\ and \Oone\ emissions are
detected.  A possible feature at 1657~\AA\ in the row 15
spectrum that could be a signature of escaping carbon atoms
\citep{Fox:1999,Cipriani:2007}, is most likely an instrumental artifact
as no emission at this wavelength appears in the row 14 spectrum.  
The background is due to grating
scattered \lya, which is variable from row to row.  Also, the spectra
have not been corrected for the odd-even row variation noted in
Section~\ref{cal} which is estimated to be $\sim$25\% for \lyb\ and
$\sim$10\% for \Oone\ \lam 1304, based on the disk observations discussed
in Section~\ref{dayglow}.

\section{Discussion}

\subsection{Exospheric hydrogen model}
\label{hmodel}

For the analysis of the \lyb\ profile we follow the same procedure as
\citet{Anderson:1971}, using the exospheric model of
\citet{Chamberlain:1963} but ignoring satellite orbits, as they can be
excluded by the observed \lyb\ brightness near 30,000~km planetocentric
distance.  For solar minimum conditions we take the exobase to be at
200~km and the exobase temperature, $T_e$, to be
200~K \citep{Krasnopolsky:2002b,Fox:2009}, leaving the hydrogen density
at this level as a variable.  We assume that \lyb\ is optically thin
and calculate a fluorescence efficiency (g-factor) of $3.9 \times 10^{-6}$
photons~s$^{-1}$~atom$^{-1}$ at 1~AU using a solar
minimum line profile and flux from \citet{Lemaire:2002}.  The interplanetary
\lyb\ background is fixed at 1.0~rayleigh based on the subsequent Jupiter
observations that were used to derive the detector flat-field at 1026~\AA\
(see Section~\ref{cal}).  All fluxes are referenced to row 15 which is
the nominal Alice boresight and which is used for almost all of the
stellar calibration measurements.

The result is shown by the solid curve in Fig.~\ref{lb_offset}.  The
derived H density at 200~km is $2.5 \times 10^5$~cm$^{-3}$ and the
escape flux is $7.8 \times 10^7$~cm$^{-2}$~s$^{-1}$.  The model
atmosphere is given in Table~\ref{denstable}.  Optical depth
unity along the line-of-sight is at $\sim$5,000~km projected
planetocentric distance.  From the radiative transfer model of
\citet{Anderson:1971}, applied to the {\it Mariner} \lya\ data,
we expect the single scattering intensity to be about a factor of two
higher than the radiative transfer corrected intensity at this optical 
depth, and this is consistent with the data shown in Fig.~\ref{lb_offset}.
\citeauthor{Anderson:1971}, for their solar maximum
observations, assuming $T_e = 350$~K at an exobase altitude of 250~km,
found an H density and escape flux of $3.0 \times
10^4$~cm$^{-3}$ and $1.8 \times 10^8$~cm$^{-2}$~s$^{-1}$,
respectively.  Fig.~\ref{lb_offset} also shows a model for 260~K
(dashed line), which, normalized at 30,000~km, provides what
superficially appears to be a better fit to the observed profile.
While it is difficult to choose between the models for planetocentric
distances greater than 10,000~km, the absence of a strong optical
depth effect in the latter suggests that the lower temperature model,
corresponding to a typical exospheric temperature at solar minimum,
is probably correct.  For 260~K, the H density and escape flux are $9.0
\times 10^4$~cm$^{-3}$ and $1.34 \times 10^8$~cm$^{-2}$~s$^{-1}$,
respectively.

The same models, applied to \lya\ using a
\lya/\lyb\ ratio of 250 derived from the IPM measurements, are shown in
Fig.~\ref{la_offset}.  The agreement is excellent and the deviation
from the optically thin emission is consistent with an optical depth
along the line-of-sight of 1 at 10,000~km planetocentric distance.
There is no apparent need for a suprathermal H component as suggested
by \citet{Chaufray:2008} and \citet{Clarke:2009}.

An interesting measurement of exospheric \lya\ emission from {\it Mars
Express} has recently been reported by \citet{Galli:2006}.  They found
that the Neutral Particle Detector of the ASPERA-3 experiment was
sensitive to \lya\ photons and measured a signal, attributed to
exospheric hydrogen, out to a tangent height of 7,250~km above the
Martian limb.  Considering their large measurement uncertainties that
include calibration, statistics, pointing, and background subtraction,
their measured emission profile is in general accord with the Alice
data shown in Fig.~\ref{la_offset}.  However, they interpret this
profile in terms of an optically thin resonance scattering model and
derive an apparent temperature $> 600$K.  As noted above, we find that
the \lya\ emission is optically thick below 10,000~km planetocentric
distance ($\sim$6,600~km above the limb), which leads to a flatter
spatial distribution and consequently the appearance of a higher
than actual exospheric temperature.

\subsection{Two-component oxygen model}

For oxygen we use a two-component model, again taking for the cold
component, $T_e$, to be 200~K, and an oxygen density at 200 km of $3.0
\times 10^7$~cm$^{-3}$ \citep{Fox:2009}.  The curve in Fig.~\ref{oalt}
represents an added hot component of 1200~K with an oxygen density at
200 km of $1.0 \times 10^5$~cm$^{-3}$ (see Table~\ref{denstable}).  The
density decrease with altitude is considerably faster than the
predictions of recent exospheric models of \citet{Chaufray:2009} and
\citet{Valeille:2010a} based on a hot atomic O source due to
dissociative recombination of O$_2^+$, and which are necessary to
support a stoichiometric escape of water vapor from the atmosphere of
Mars.  However, it has been noted that such inferences from a single
viewing geometry during a period of low solar activity can be
misleading as this mechanism is quite sensitive to solar activity and
is dependent on solar zenith angle at the observation point.

Nevertheless, the present observations raise concern about some of
the assumptions and physical parameters used in the recent modeling of
oxygen escape from Mars.  \citet{Fox:2009}, in their comparison of
exobase and Monte Carlo models, summarize the literature on modeling
efforts from the past few decades and conclude that ``efforts to
balance the escape rates in the stoichiometric proportion of water are
premature.''  \citeauthor{Fox:2009} focus on a comparison of escape
rates and therefore do not compute the oxygen density profiles from
their models.  Such calculation is warranted by the present data
which would allow for a determination of the line-of-sight column
densities appropriate to the Alice observations.

Similar questions arise in the analogous modeling of the
hot oxygen environment around Venus \citep{Groller:2010}.
\citet{Bovino:2011} discuss the need for accurate data on energy transfer
collisions between hot oxygen atoms and the neutral atmosphere (they
are mainly interested in helium) which is at the core of the escape
models.  Finally, we note that \citet{Simon:2009} point out that
an additional constraint on the models might be provided by SPICAM
measurements of the forbidden \Oone\ $\lambda$2972 ($^1$S -- $^3$P) line,
which is produced by both photodissociation of \cotwo\ and by dissociative
recombination of O$_2^+$.

\section{Conclusion}

The \rosetta\ swing-by of Mars on 25 February 2007, provided the first
spectroscopic observations of exospheric hydrogen and oxygen on Mars from
outside Mars' atmosphere since the {\it Mariner 6} and {\it 7} fly-bys in 1969.  
The spatial distribution of
\Hone\ \lya\ out to beyond 30,000 km from the planet's center
is similar to that found from {\it Mariner 6} and {\it 7}.  A Chamberlain model,
with current solar minimum model values of exospheric temperature of
200~K and an exobase altitude of 200~km, provides a good fit to the
observed \lyb\ profile, although the data do not exclude
temperatures up to $\sim$260~K.  A suprathermal component, suggested by
several authors, is not needed to match the data.  The  
hydrogen escape flux derived from the 200~K model, $7.8 \times 
10^7$~cm$^{-2}$~s$^{-1}$, is comparable to that derived from the earlier
measurements.  The distribution of atomic oxygen, derived from
1304~\AA\ emission observed to altitudes of 1000~km above the limb, is
not consistent with recent models of oxygen escape based on the
production of suprathermal oxygen atoms by dissociative
recombination of O$_2^+$.

\begin{center}{\bf Acknowledgments}\end{center}

We thank the ESA Rosetta Science Operations Centre (RSOC) and Mission
Operations Center (RMOC) teams for their expert and dedicated help in
planning and executing the Alice observations of Mars.  We thank
Darrell Strobel for helpful discussions.  The Alice team acknowledges
continuing support from NASA's Jet Propulsion Laboratory through
contract 1336850 to the Southwest Research Institute.  The work at
Johns Hopkins University was supported by a sub-contract from Southwest
Research Institute.

\clearpage
\bibliographystyle{model2-names}







\renewcommand\baselinestretch{1.2}%

\clearpage

\begin{table*}
\begin{center}
\caption{Observation Parameters.  \label{obstable}}
\medskip
\begin{tabular}{@{}lccc@{}}
\hline
Observation ID  & AL03 & AL10E & AL11B \\
\hline
Start time (UT 2007) & 24 Feb 18:28:14 & 24 Feb 20:13:14 & 25 Feb 03:33:02 \\
Boresight pointing & disk center & offset 2.5\arcdeg, 7.5\arcdeg\ W & E limb scan \\
Distance to Mars$^a$ (km) & 239,800 & 184,200 & 51,600 \\
Solar elongation$^a$ & 164.3\arcdeg\ & 164.0\arcdeg\ & 24.3\arcdeg\ \\
Longitude of tangent point at limb$^b$ & 213.9\arcdeg\ & 149.7\arcdeg\ & 263.6\arcdeg\ \\
Latitude of tangent point at limb$^b$ & --1.8\arcdeg\ & --4.1\arcdeg\ & --26.5\arcdeg\ \\
Solar zenith angle at tangent point$^b$ & 15.7\arcdeg\ & 74.3\arcdeg\ & 67.9\arcdeg\ \\
Data mode & Histogram & Histogram & Pixel list \\
Total integration time (s) & 4112 & 483 + 1028 & \\
\hline
\multicolumn{4}{l}{$^a$At start of observation.} \\
\multicolumn{4}{l}{$^b$At sub-observer point for AL03.} \\
\end{tabular}
\end{center}
\end{table*}

\clearpage

\begin{table*}
\begin{center}
\caption{Model Atmosphere Densities.  \label{denstable}}
\medskip
\begin{tabular}{@{}cccc@{}}
\hline
Altitude & H & O (cold) & O (hot) \\
 (km) & (cm$^{-3}$) & (cm$^{-3}$) & (cm$^{-3}$) \\
\hline
\hline
  200  &   2.50 $\times$ $10^5$ & 3.00 $\times$ $10^7$ & 1.00 $\times$ $10^5$ \\
  400  &   1.69 $\times$ $10^5$ & 6.86 $\times$ $10^4$ & 3.70 $\times$ $10^4$ \\
  600  &    1.19 $\times$ $10^5$ & 288.0 &  1.49 $\times$ $10^4$ \\
  800  &    8.61 $\times$ $10^4$ & 2.04 & 6520. \\
  1000 &     6.42 $\times$ $10^4$ & 0.023 & 3080. \\
  1200 &     4.90 $\times$ $10^4$ & 3.7 $\times$ $10^{-4}$ & 1550. \\
  1400 &     3.82 $\times$ $10^4$ &  & 830. \\
  1600 &     3.03 $\times$ $10^4$ &  &  \\
  1800 &     2.44 $\times$ $10^4$ &  &  \\
  2000 &     1.99 $\times$ $10^4$ &  &  \\
\hline
  4000  &    4330. &  &  \\
  6000  &    1590. &  &  \\
  8000  &    757. &  &  \\
  10000 &     422. &  &  \\
\hline
  15000  &    144. &  &  \\
  20000  &    67.5 &  &  \\
  25000  &    37.5 &  &  \\
  30000  &    23.2 &  &  \\
  35000  &    15.5 &  &  \\
\hline
\end{tabular}
\end{center}
\end{table*}

\renewcommand\baselinestretch{1.6}%

\clearpage 
\begin{center}{\bf FIGURE CAPTIONS}\end{center}

\noindent
Fig.~\ref{offset}.  Projection of the Alice slit on the sky with the common
boresight offset 2.5\arcdeg\ from the center of Mars along the
equator.  
Orange grid lines outline the illuminated region of the disk.
The shape of the multi-segment slit is shown.  Detector row
numbers increase from left to right with the boresight (+) in row 15.

\vspace*{0.1in}
\noindent
Fig.~\ref{limb}.  Projection of the Alice slit on Mars at the beginning of the
limb scans following closest approach.  
Orange grid lines outline the illuminated crescent of the disk.
Detector row numbers increase
from right to left with the boresight (+) in row 15.

\vspace*{0.1in}
\noindent
Fig.~\ref{limb_time}.  Projection of individual rows of the Alice slit
above the Mars limb during the slow limb scan following closest
approach.  Initially, each spatial pixel projected to 280~km in
altitude but as Rosetta receded from Mars the projected size of each
pixel increased.  The scan also slowly shifted the boresight
$\sim$250~km towards Mars over a 15-minute period.  The cross-hatched
areas indicate the time during which photon events were accumulated
while the horizontal lines show the times for which photon events were
co-added for each row.

\vspace*{0.1in}
\noindent
Fig.~\ref{day_spec}.  The dayglow spectrum of Mars is a composite of the five
central rows of four exposures, each of 1028 seconds. For comparison
with previous work, we also show the Mars full disk spectrum recorded
by the Hopkins Ultraviolet Telescope (HUT) at 4~\AA\ spectral resolution
in March 1995 (a full solar
cycle earlier) \citep{Feldman:2000}, convolved to the spectral
resolution of Alice and multiplied by a factor of 0.8.

\vspace*{0.1in}
\noindent
Fig.~\ref{lb_offset}.  \Hone\ \lyb\ was detected along the full length
of the Alice slit for both offsets. The data are shown as a histogram:
black, 2.5\arcdeg\ offset, red:  7.5\arcdeg\ offset. The errors shown
are statistical in the count rate.  Optically thin Chamberlain models
without satellite orbits for T=200 K (solid line) and 260 K (dashed line)
are shown, superimposed on a 1~rayleigh interplanetary background.
The radius of Mars is indicated.

\vspace*{0.1in}
\noindent
Fig.~\ref{la_offset}.  Same as Fig.~\ref{lb_offset} for \Hone\ \lya.
The shape of this profile is very similar to the slant intensity
profiles derived from the {\it Mariner 6} and {\it 7} fly-bys by
\citet{Barth:1971}, although lower in absolute brightness as the
{\it Mariner} fly-bys occurred at a time of high solar activity.  The same
models are shown, superimposed on an interplanetary background of 250
rayleighs.

\vspace*{0.1in}
\noindent
Fig.~\ref{limbspec}.  Extracted limb scan spectra.  For row 15 (top),
the photon counts from the first 12 exposures (376~s, see
Fig.~\ref{limb_time}) were co-added.  For row 16 (bottom), the last 14
exposures (476~s) were co-added.  Only \Hone\ and \Oone\ emissions are
detected.

\vspace*{0.1in}
\noindent
Fig.~\ref{oalt}.  Extracted \Oone\ $\lambda$1304 brightness from rows 14--17
as a function of altitude above the Martian limb.  The vertical bars
illustrate the extent of the trapezoidal altitude weighting function
for each row while the horizontal bars are the statistical uncertainty
in the count rate.  The two-component oxygen model shown is described in the
text.

\renewcommand\baselinestretch{1.2}%
\setcounter{figure}{0}

\begin{figure*}[ht]
\begin{center}
\includegraphics*[width=\textwidth,angle=0.]{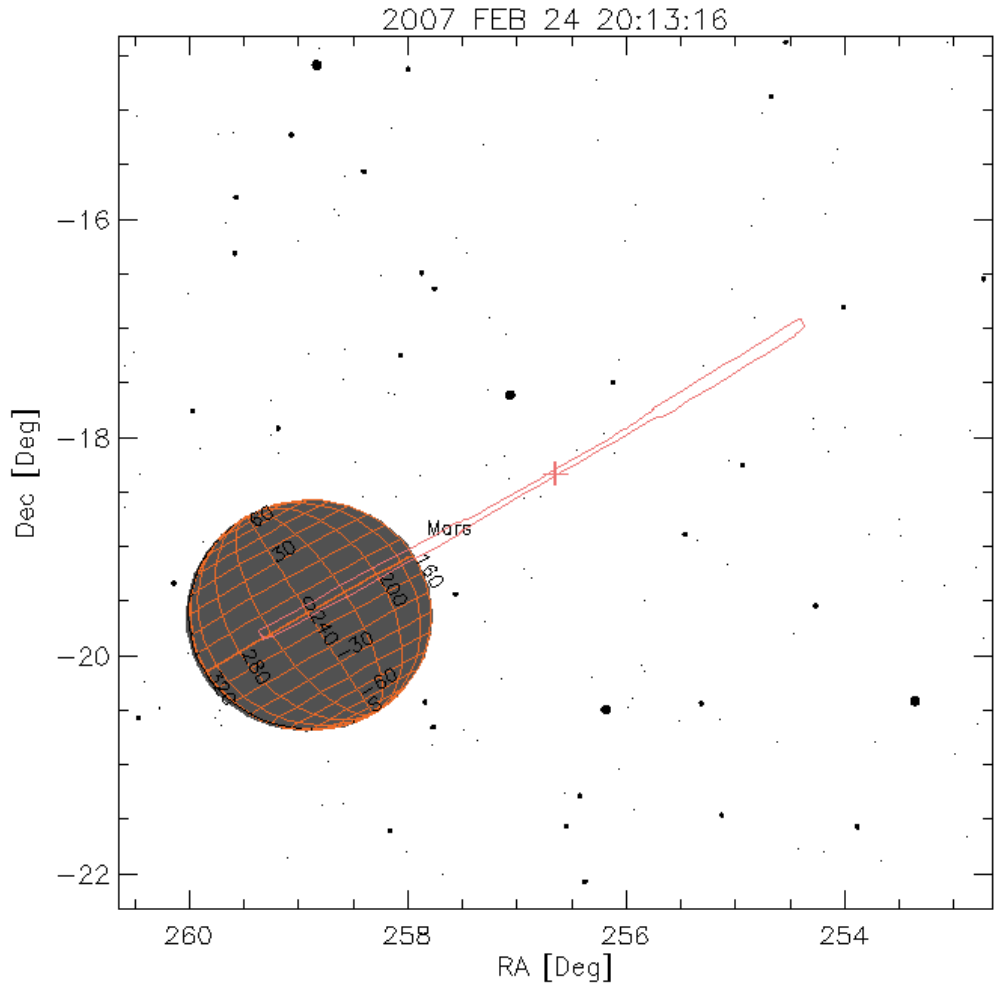}
\caption[]{Projection of the Alice slit on the sky with the common
boresight offset 2.5\arcdeg\ from the center of Mars along the
equator.  Orange grid lines outline the illuminated region of the
disk.  The shape of the multi-segment slit is shown.  Detector row
numbers increase from left to right with the boresight (+) in row 15.
\label{offset} }
\end{center}
\end{figure*} 

\begin{figure*}[ht]
\begin{center}
\includegraphics*[width=\textwidth,angle=0.]{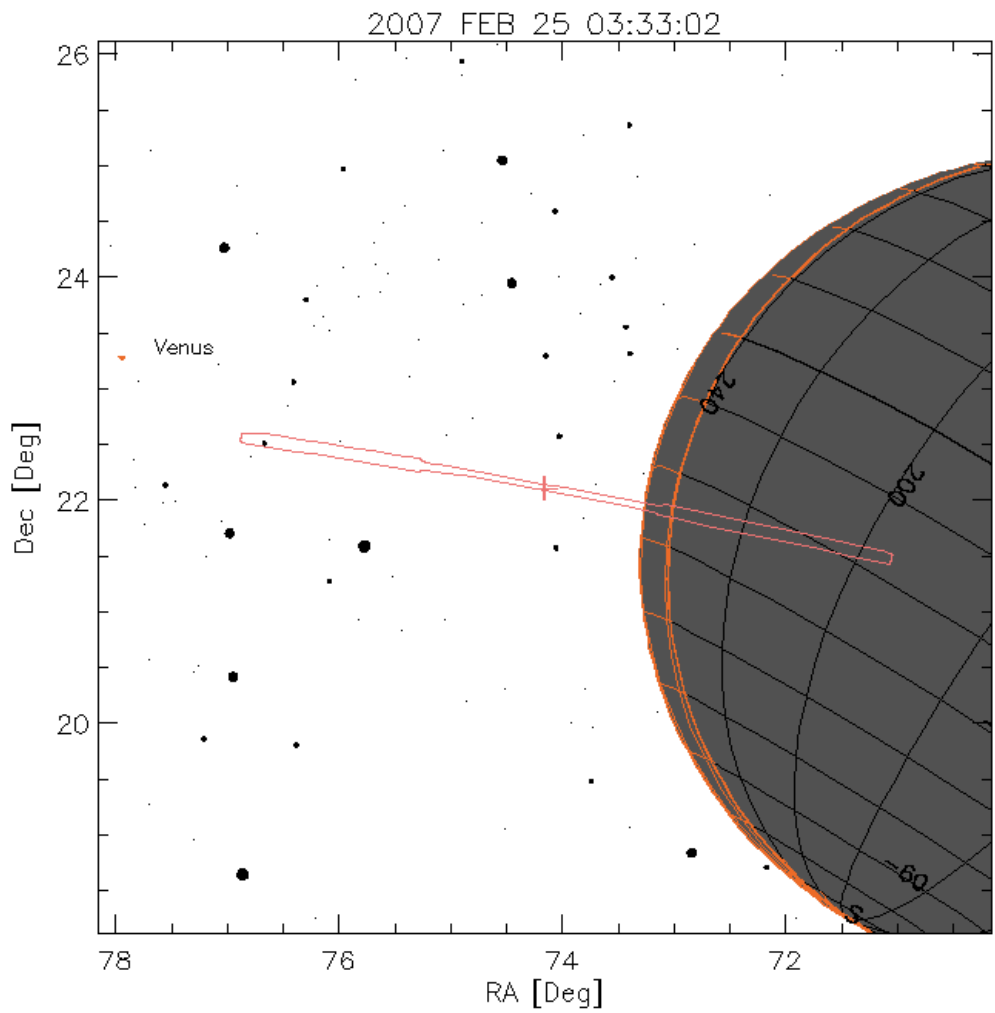}
\caption[]{Projection of the Alice slit on Mars at the beginning of the
limb scans following closest approach.  Orange grid lines outline the
illuminated crescent of the disk.  Detector row numbers increase from
right to left with the boresight (+) in row 15.  \label{limb} }
\end{center}
\end{figure*} 

\begin{figure*}[ht]
\begin{center}
\includegraphics*[width=\textwidth,angle=0.]{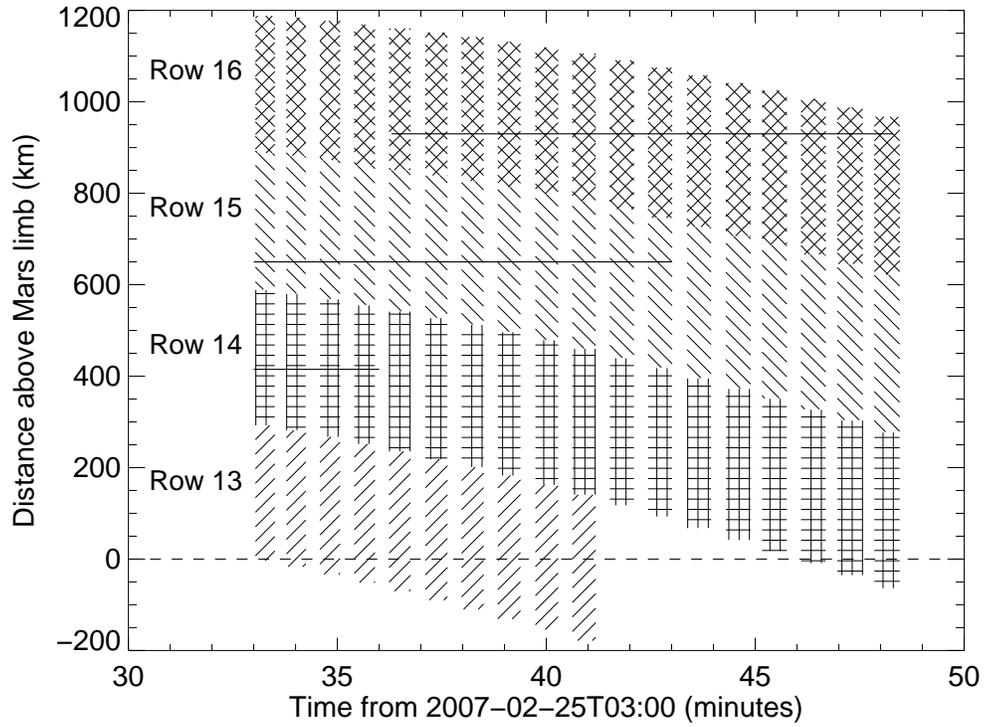}
\end{center}
\caption[]{Projection of individual rows of the Alice slit above the
Mars limb during the slow limb scan following closest approach.  
Initially, each spatial pixel projected to 280~km in altitude but as
Rosetta receded from Mars the projected size of each pixel increased.
The scan also slowly shifted the boresight $\sim$250~km towards Mars
over a 15-minute period.  The cross-hatched areas indicate the time
during which photon events were accumulated while the horizontal lines
show the times for which photon events were co-added for each row.
\label{limb_time} }
\end{figure*} 

\begin{figure*}[ht]
\begin{center}
\includegraphics*[width=\textwidth,angle=0.]{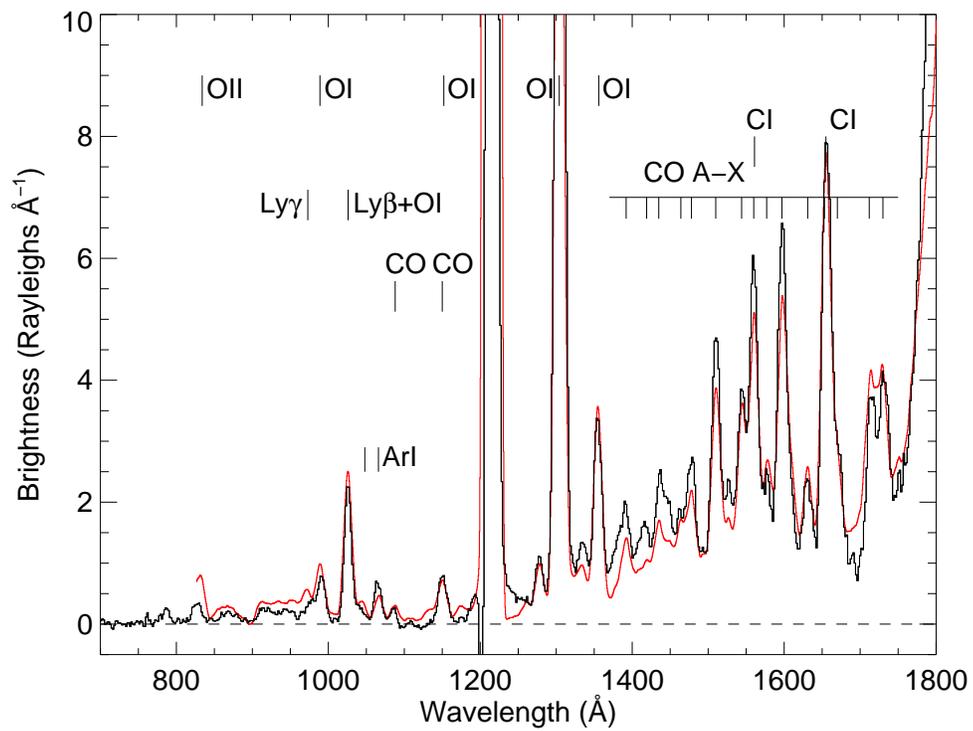}
\caption[]{The dayglow spectrum of Mars is a composite of the five
central rows of four exposures, each of 1028 seconds. For comparison
with previous work, we also show the Mars full disk spectrum recorded
by the Hopkins Ultraviolet Telescope (HUT) at 4~\AA\ spectral resolution
in March 1995 (a full solar
cycle earlier) \citep{Feldman:2000}, convolved to the spectral
resolution of Alice and multiplied by a factor of 0.8.  \label{day_spec} }
\end{center}
\end{figure*}

\begin{figure*}[ht]
\begin{center}
\includegraphics*[width=\textwidth,angle=0.,clip]{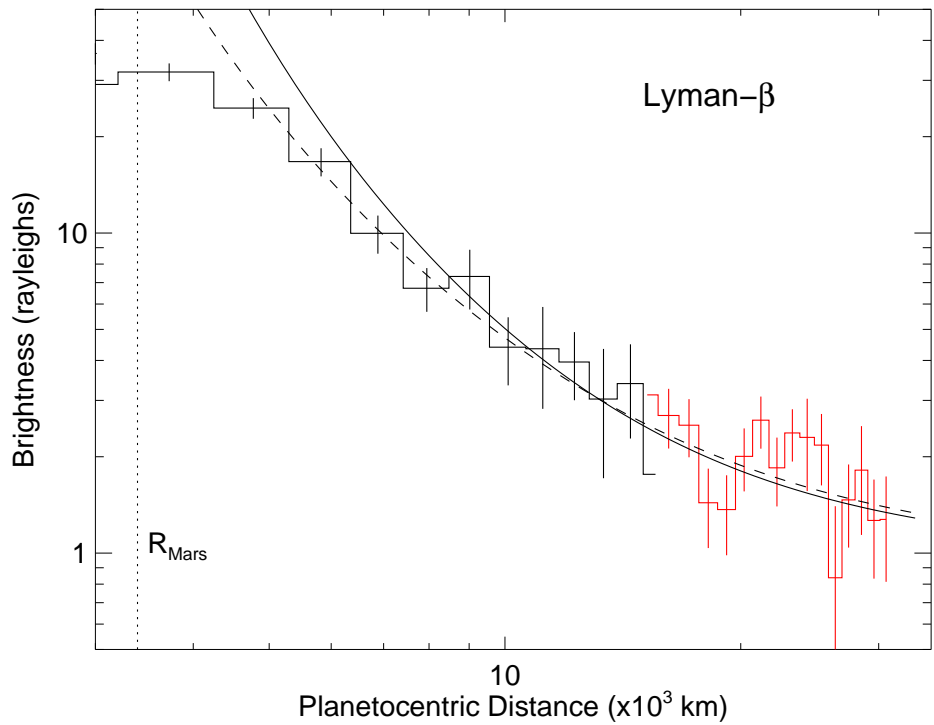}
\caption[]{\Hone\ \lyb\ was detected along the full length of the Alice
slit for both offsets. The data are shown as a histogram: black,
2.5\arcdeg\ offset, red:  7.5\arcdeg\ offset. The errors shown are
statistical in the count rate.  Optically thin Chamberlain models
without satellite orbits for T=200 K (solid line) and 260 K (dashed
line) are shown, superimposed on a 1~rayleigh interplanetary
background.  The radius of Mars is indicated.  \label{lb_offset} }
\end{center}
\end{figure*} 

\begin{figure*}[ht]
\begin{center}
\includegraphics*[width=\textwidth,angle=0.,clip]{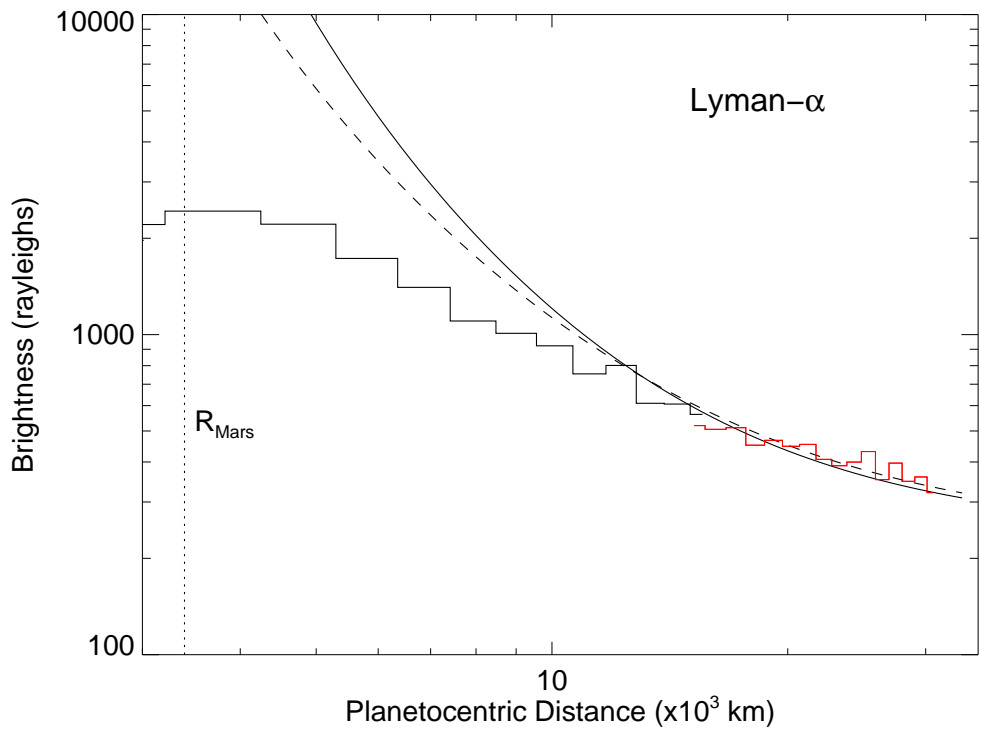}
\caption[]{Same as Fig.~\ref{lb_offset} for \Hone\ \lya.  The shape of
this profile is very similar to the slant intensity profiles derived
from the {\it Mariner 6} and {\it 7} fly-bys by \citet{Barth:1971}, although lower
in absolute brightness as the {\it Mariner} fly-bys occurred at a time of
high solar activity.  The same models are shown, superimposed on an
interplanetary background of 250 rayleighs.  \label{la_offset} }
\end{center}
\end{figure*} 

\begin{figure*}[ht]
\begin{center}
\includegraphics*[width=0.95\textwidth,angle=0.]{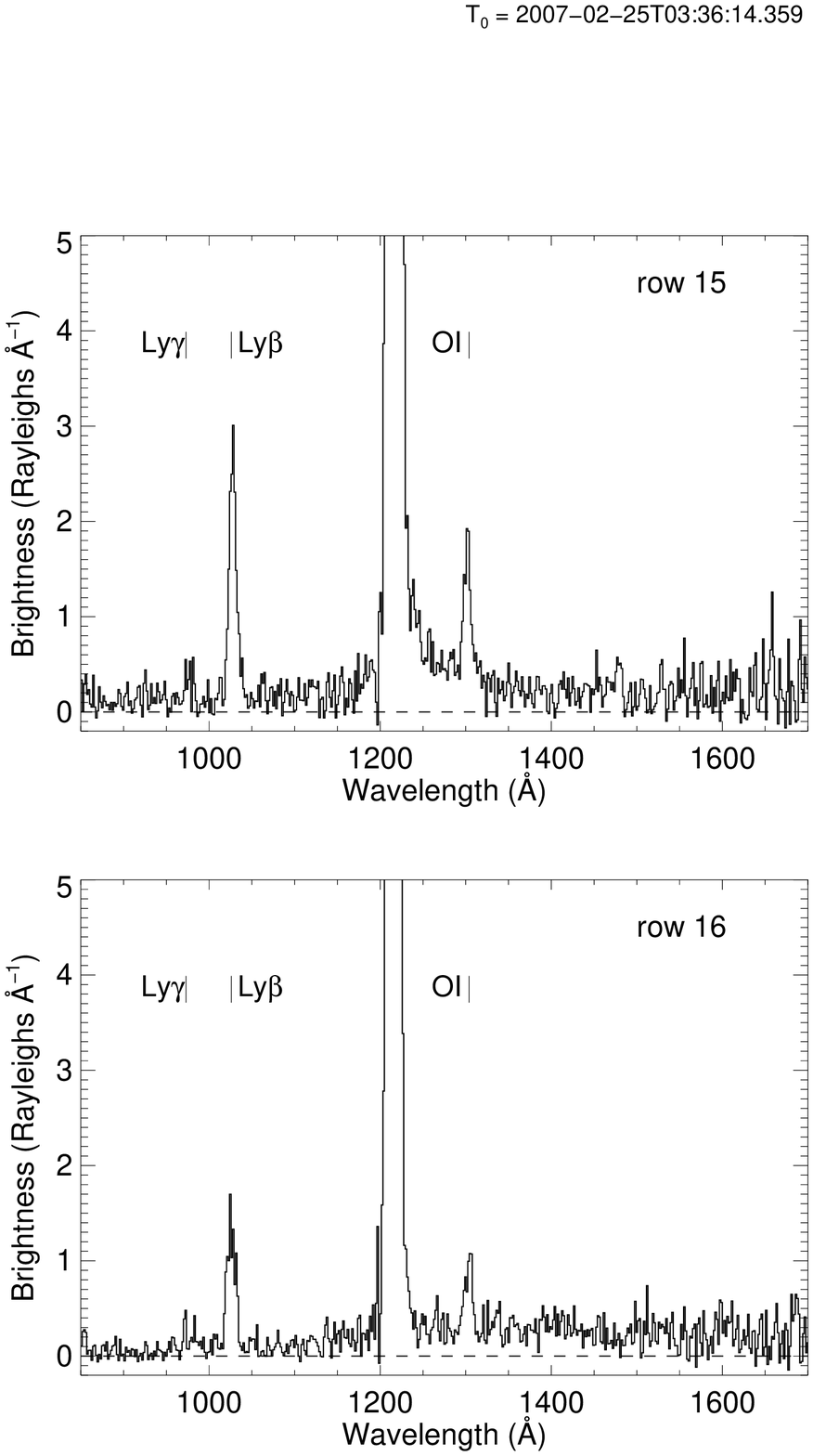}
\vspace*{-0.2in}
\caption[]{Extracted limb scan spectra.  For row 15 (top), the photon
counts from the first 12 exposures (376~s, see Fig.~\ref{limb_time})
were co-added.  For row 16 (bottom), the last 14 exposures (476~s) were
co-added.  Only \Hone\ and \Oone\ emissions are detected.  \label{limbspec} }
\end{center}
\end{figure*} 

\begin{figure*}[ht]
\begin{center}
\includegraphics*[width=1.0\textwidth,angle=0.]{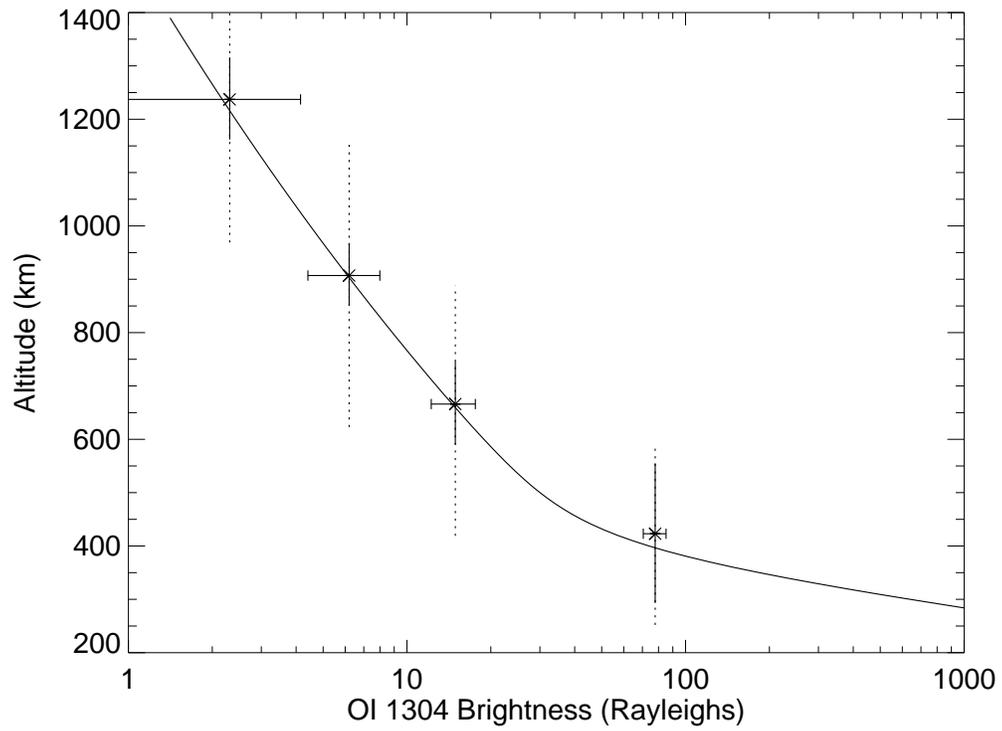}
\caption[]{Extracted \Oone\ $\lambda$1304 brightness from rows 14--17
as a function of altitude above the Martian limb.  The vertical bars
illustrate the extent of the trapezoidal altitude weighting function
for each row while the horizontal bars are the statistical uncertainty
in the count rate.  The two-component oxygen model shown is described in the
text.  \label{oalt} }
\end{center}
\end{figure*}

\end{document}